\documentclass[proof]{pasj00}

\begin{document}
\SetRunningHead{M. Ishiguro}{Scattered Light Correction of HAYABUSA/AMICA Data}
\Received{2013/10/29}
\Accepted{2014/01/17}

\title{Scattered Light Correction of HAYABUSA/AMICA Data and Quantitative
Spectral Comparisons of Itokawa}

\author{Masateru \textsc{Ishiguro} %
  \thanks{Visiting scientist at the Department of Earth and Space
  Sciences, University of California at Los Angeles, 595 Charles Young
  Drive East, Los Angeles, CA  90095-1567, USA}}
\affil{Department of Physics and Astronomy, Seoul National University,\\
Gwanak, Seoul 151-742, Republic of Korea}
\email{ishiguro@astro.snu.ac.kr}


\maketitle

\begin{abstract}
The Hayabusa spacecraft rendezvoused with its target asteroid
25143 Itokawa in 2005 and brought an asteroidal sample back to the Earth in
2010. The onboard camera, AMICA, took more than 1400 images of Itokawa
during the rendezvous phase. It was reported that the AMICA images were
severely contaminated by light scatter inside the optics. The effect
made it difficult to produce the color maps at longer
wavelengths ($>$800 nm). In this paper, we demonstrate a method to
subtract the scattered light by investigating the dim halos of Itokawa and
the Moon taken by AMICA during the inflight operation. 
As the result, we found that the overall data reduction
scheme including the scattered light correction enables to recognize
~3\% regional differences in the 
relative reflectance spectra of Itokawa.
We confirmed that the color variation in Itokawa was largely attributed to
space weathering.
\end{abstract}

\section{Introduction}
Multiband astronomical photometry is powerful technique that is used to derive the magnitudes
and colors of celestial bodies. In solar system astronomy, taxonomic
types of small bodies are often derived based on multiband photometry
{e.g. \cite{Hasegawa2014,Takahashi2014})
or spectroscopy (e.g. \cite{Kuroda2014}).
The data is typically acquired at dark locations without a
significant amount of light contamination. The impurities of unwanted light such as the
background airglow emissions have been excluded by several techniques,
including aperture photometry. Conversely, remote-sensing
observations of solar system objects are pivotal for studying the
reflectances of extended objects in order to determine the regional diversities of
color on the target bodies, which helps us to hypothesize about the formation
and alternation processes of the objects. However, lights from the
sun and their target bodies can be an inherent problem in space
exploration data, which is different from the problems faced with
astronomical observations near a light source.

It is reported that approximately 10\% of light inside astronomical
instruments, hereafter called 'scattered light', contains contaminated
data taken by onboard cameras for various space explorations (NEAR/MSI~\cite{Murchie2002},
Galileo/SSI~\cite{Klaasen1997}, Clementine/UVVIS~\cite{Pieters1994}).
Since the brightness of the scattered light could
depend on the wavelengths and positions of the objects inside the detectors, it could
skew the surface brightness distribution and eventually the reflectance and
color maps of the target bodies. This effect would be worse for a small dark
object in a bright terrain since the dark objects are subject to
light leakage from adjacent brighter areas. For the case of the Clementine
UVVIS, the flux calibration method exhibited a serious problem since the
reference site used for the calibration, i.e., the Apollo landing site, was located
on a dark mare near the bright highland massifs
\citep{Robinson2001,Robinson2003}. During the NEAR mission, the multispectral
imager, MSI, was contaminated by a portion of the burn products that occurred during a
spacecraft anomaly. This caused a significant amount of light to be scattered
\citep{Murchie2002}. Meanwhile, \citet{Gaddis1995} thoroughly examined the problem in the Galileo
solid-state imaging subsystem (SSI) images, in which they were able to
distinguish between the effects of scattered light, which resulted from
internal scattering, and those of stray light, which is light outside of the
optics. They were able to remove the scattered light from the obtained images
using an attenuation function. In short, the scattered light analysis is
crucially important and inevitable for the imaging data taken by a
spacecraft's onboard cameras.

The asteroid multiband imaging camera (AMICA) is an onboard device of
the Hayabusa spacecraft. It is the first Japanese fully-fledged
multiband imaging 
device to explore interplanetary bodies. AMICA took more than 1400
images of its mission target asteroid, Itokawa, between September and
November 2005. These images revealed the nature of Itokawa, including its rubble-pile
structure and space weathering effects on its surface
\citep{Fujiwara2006,Demura2006,Saito2006,Yano2006,Abe2006a}. AMICA is equipped with
bandpass
filters similar to those of the Eight Color Asteroid Survey (ECAS),
which have been applied for astronomical observations of asteroids
\citep{Tedesco1982,Zellner1985}. The AMICA effective wavelengths with
respect to the Sun are the ul-band (381 nm), b-band (429 nm), v-band
(553 nm), w-band (700 nm), x-band (861 nm), p-band (960 nm) and zs-band
(1008 nm) \citep{Ishiguro2010}. Early analysis of the AMICA
data revealed that Itokawa exhibits a large variation of reflectance and color
\citep{Saito2006}. Later, \citet{Ishiguro2007} derived the color map of
Itokawa using AMICA's b-, v-, and w-band images, and determined the
space weathering effects. However, no data at longer wavelengths, i.e., x-,
p-, and zs-bands, have yet to be used due to a problem discussed below.

During the Earth swing-by phase in 2004, we took images of the Moon and
noticed an unexpected broadening of the light at longer channels
\citep{Ishiguro2010}. The observed light consisted of not only the focused
images of the Moon, but also a faint glow adjacent to the lunar
limb. The sky close to the objects was therefore brighter than the bias plus
dark level. Herein, we refer to the glow as 'dim halo'. The dim halo was
very pronounced in near-infrared channels, i.e. in the x-, p- and zs-bands, but only
marginally noticeable in the ul-band. Later, we noticed that the dim halo
was more pronounced in the Itokawa's
images taken during the rendezvous phase. We found that the dim halos were
caused by the scattering inside of the AMICA optics, as seen in the
previous mission imaging data and discussed in Section 2-1. Since no
method was developed to subtract the scattered light components in AMICA
images thus far, we could not provide color maps of Itokawa using channels
other than b-, v-, and w-bands. In particular, the p-band data were expected
to provide crucial results of the space weathering and mineralogy on Itokawa
because the filter was designed to obtain images in the center of pyroxene
absorption band.

In this paper, we provide a technical description on the subtraction
of scattered light in AMICA images. By investigating
the dim halos of the Moon and Itokawa,
we obtained widespread
attenuation functions of incident light for all AMICA channels during the
scattered-light correction. We subtracted the scattered light from the
original images using the attenuation function and demonstrated the effect
this method has on the AMICA images. Using the technique described herein,
we provide tentative results of color maps and discuss the
association between the color variation and space weathering effects.

\section{Methodology}

In this chapter,
we first outline the evidence for scattered
light in AMICA images, and then provide a correction method.
Finally, we evaluate the effects of the correction using shadowed
areas on Itokawa. All data were obtained from a public data archive, i.e., the
{\it Hayabusa Project Science Data Archive}.
\footnote{http://darts.isas.jaxa.jp/planet/project/hayabusa/}

\subsection{Observed Evidences}
Figure~\ref{fig:f1} shows a p-band image of Itokawa. To emphasize
a dim halo, we drew the contours for the 23--230 data numbers (DNs), which
corresponds to roughly 1\%--10\% of the average Itokawa surface
brightness, i.e., a 2300 DN. The intensity of the dim halo decreased
nearly concentrically as the distance from Itokawa increased. In
addition, the intensity of the dim halo was wavelength dependent on the
distance from Itokawa. Figure~\ref{fig:f2} shows the normalized
intensity of a dim halo from the Itokawa images. The intensity 
of the dim halo was brightest at the longest wavelength (zs-band) and
dimmer at shorter wavelengths. The intensity was lowest in the b-, v-, and
w-bands, but slightly higher at the shortest wavelength, i.e., the ul-band.

\begin{figure}
 \begin{center}
  \includegraphics[width=80mm]{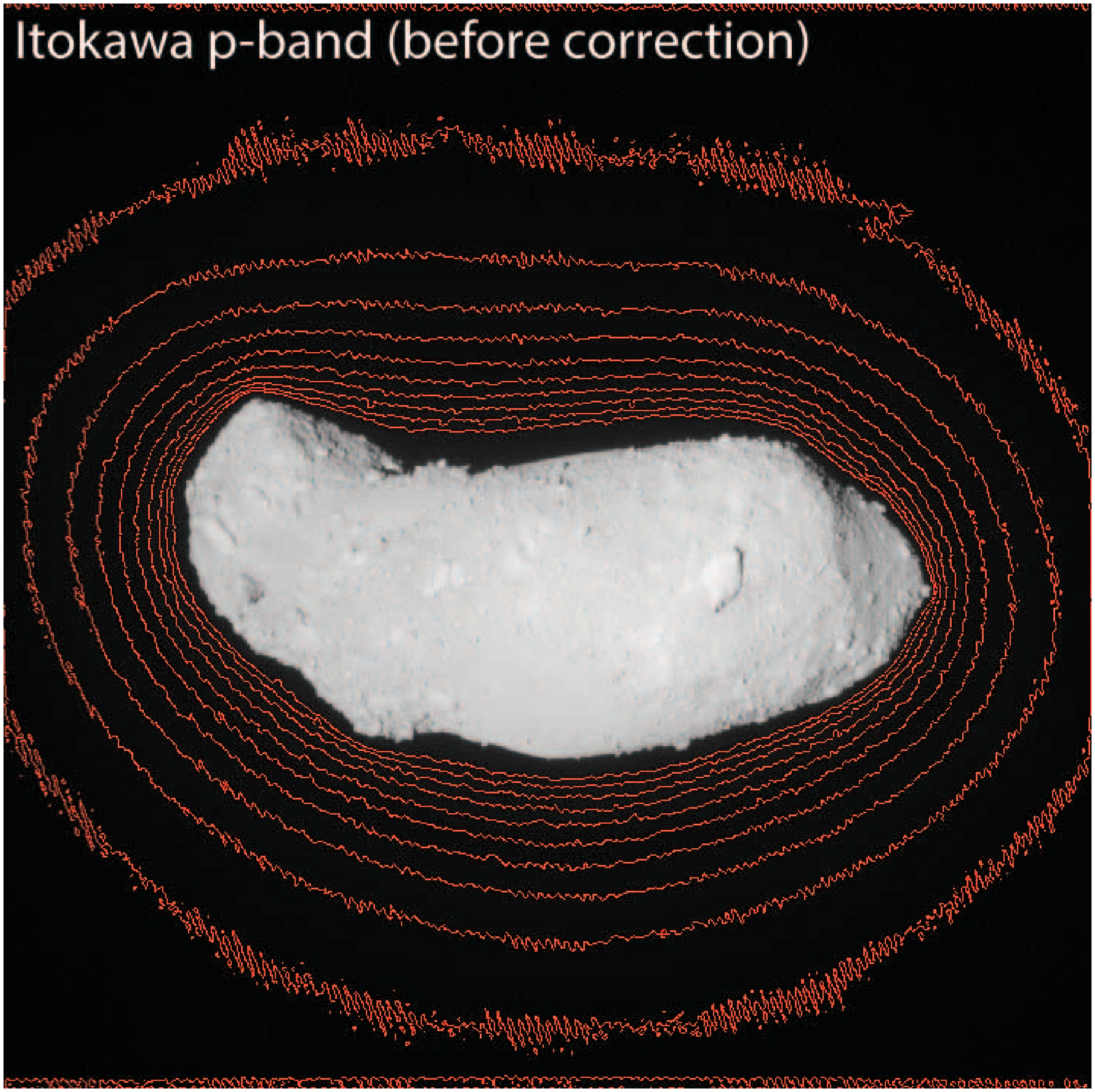}
  \includegraphics[width=80mm]{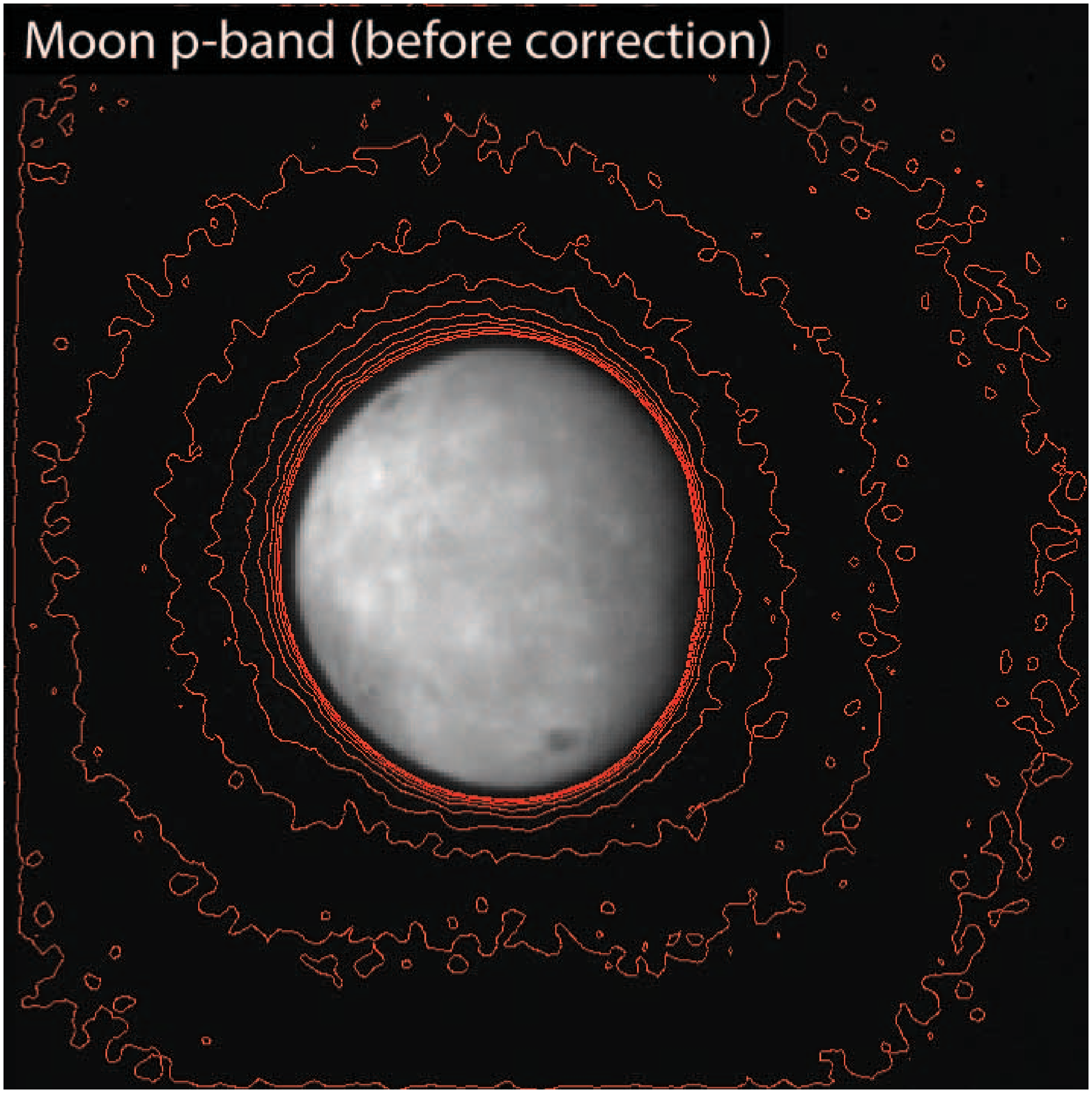}
  \caption{Images before the scattered light
  correction. (left) The p-band full-size (1024$\times$1024 pixels) image
  of   Itokawa taken at 7.5 km from the Itokawa's  surface (File name:
  ST\_2418807291). (right) The p-band partial area (256$\times$256
  pixels) image of the moon taken during the Earth swing-by (File name:
  ST\_1035463706). In both images, bias, readout smear, and flatfield
  were corrected in a manner written in \citet{Ishiguro2010}. The
  counter lines corresponding to 1\% -- 10\% of Itokawa's disk intensity
  at 1\% interval were drawn to bring out a hidden dim halo in the
  image.} \label{fig:f1} 
 \end{center}
\end{figure}

\begin{figure}
  \begin{center}
   \includegraphics[width=80mm]{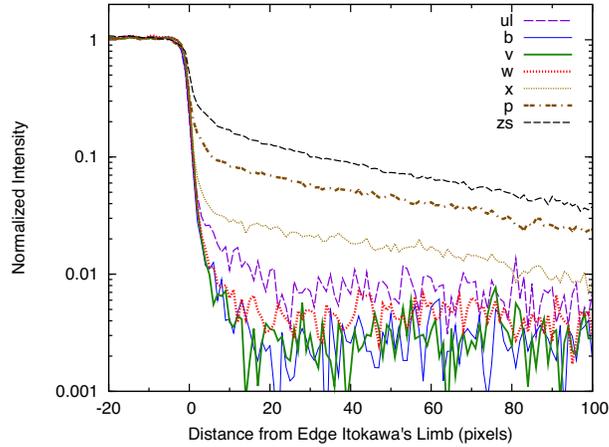}
   \caption{Intensities of Itokawa's dim halo at different
   wavelengths. These images were taken 13.9~km from the Itokawa
   surface on September  25, 2005 (File name: ST\_2406708286 --
   ST\_2406766067). These profiles were normalized to the intensity of the
   Itokawa disk, i.e., an intensity of 1. The intensities dropped to $<$1\% level in the b-, v-, and w-bands
   within a few pixels, while the intensity persisted at the 1--10\% level for the
   longer three bands, i.e., the x-, p-, and zs-bands.}\label{fig:f2}
  \end{center}
\end{figure}

We also determined that the dim halo intensity depends on the apparent size
of Itokawa. In fact, the dim halo intensity is enhanced as the apparent size of object is increased.
Figure~\ref{fig:f3} shows the dim halo intensity of Itokawa taken at a
similar rotational phase but different distance from Itokawa's surface.
The data was acquired at 7.4 and 13.9~km from the surface of Itokawa.
For convenience, herein, we call the
mission phases the Gate Position (GP) phase around 15--20~km, and the
Home Position (HP) phase around 7~km \citep{Fujiwara2006}. It is clear
that the brightness of dim halo drops rapidly when the data was taken at the
GP phase. It is likely that the intensity of dim halo may be correlated to the
total intensity of the object in the AMICA field-of-view (FOV).
On the basis of these observational evidences, it is natural to
hypothesize that the dim halo could be caused by the internal scattering of the
AMICA optical components, i.e., the baffles, lens unit, and interior wall
of the telescope tube, since the scattering properties of
these components could be wavelength-dependent. The main light source of
the scattered light could be the light from the object inside the AMICA
optics.
While the scattered light is noticeable in the dim halo, it is
natural to believe that the scattered light could be superposed on the focused
images of Itokawa and the Moon. In other words, the contamination of the
scattered light could lead to a misunderstanding of the Itokawa color maps.
Since we, the AMICA team, did not expect this effect before the
launch, we have no preflight data for the
scattered light correction. Therefore, we must establish a
subtraction method using inflight data.

\begin{figure}
  \begin{center}
  \includegraphics[width=80mm]{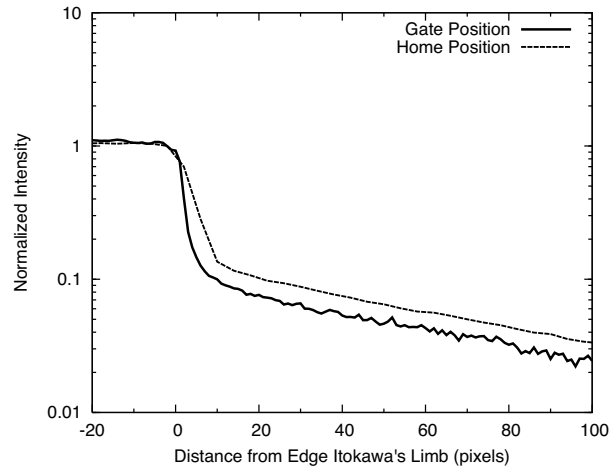}
  \caption{Comparison of the dim halo intensity taken at different
   distances. The data was acquired at the GP and
   HP phases, where the spacecraft was located 13.9 and
   7.4~km away from the Itokawa's surface (File name of
   ST\_2406766067 and ST\_2468186849), respectively.
   The distance data were taken
   based on the LIDAR measurement \citep{Abe2006b}.
}\label{fig:f3}
  \end{center}
\end{figure}

In addition to the scattered light, extra light from beyond the FOV,
which is known as 'stray light', causes additional light on detectors
\citep{Phillips2010,Gaddis1995}. For case of the AMICA, which was attached
the bottom plane of the spacecraft in a shaded area of sunshine, there
was no obvious stray light when Itokawa was in the AMICA FOV.
\citet{Ishiguro2010} examined the stray light using the data taken
at different spacecraft altitudes with respect to the sun, and
concluded that the stray light was negligible, i.e., $\ll$1 DN, in the
nominal spacecraft attitude during the GP and HP phases. For this reason, we
ignore stray light in this paper. To sum up the evidences in this
subsection, the most important process for the scattered light
correction was to derive an attenuation function that can characterize
how the incident light attenuated as a function of 
distance. Once we determine the functions, we are able to subtract the
scattered light using several methods such as image
deconvolution. Hereafter, we describe how we derived the attenuation
function using AMICA inflight data.

\subsection{Determination of AMICA Point Spread Function}

Point spread functions (PSFs) describe the two-dimensional distribution
of light emitted from an infinitely small point source. In normal
astronomical observations, this terminology is usually used for the light that
extends to tens of pixels. The AMICA PSFs near light sources,
$f_\mathrm{foc}(r)$, could be determined by observing point
sources, which were derived through the observations of stars and reported
results in the paper by \citet{Ishiguro2010}. The component forms
a focused image on the CCD detector. We empirically determined that the
following mathematical expression is suitable for characterizing the
AMICA PSFs near light sources,
\begin{equation}
f_\mathrm{foc}(r) = \exp \left( -\alpha r \right),
 \label{eq:eq1}
\end{equation}

\noindent
where $r$ is the distance in pixels from the optical center of the point
sources. The obtained parameter $\alpha$ is summarized in
Table 1.

\begin{table}
 \footnotesize
  \caption{The $\alpha$ coefficient that is used to characterize AMICA PSFs near light sources.}
\label{tab:first}
  \begin{center}
    \begin{tabular}{cccccccc}
      \hline
     filter & ul & b & v & w & x & p & zs \\
      \hline
     $\alpha$ (pixel$^{-1}$)& 1.26 & 1.28 & 1.41 & 1.85 & 1.85 & 1.60 & 1.48 \\
      \hline
    \end{tabular}
  \end{center}
\end{table}

In addition to the focused component, we considered a widespread
attenuation function as a part of the AMICA PSFs in the broad sense,
following \citet{Gaddis1995}. The attenuation function produces unfocused images on the
AMICA detector as we saw in the dim halo. We assumed that the broad PSFs,
i.e., the attenuation functions, were axisymmetric and were fitted using the
summation of Gaussian functions:

\begin{equation}
f_\mathrm{ufoc}(r) = \sum_{i=1}^N \frac{A_i}{\sqrt{2 \pi}\sigma_i} \exp
 \left(-\frac{r^2}{2 \sigma_i^2}\right),
 \label{eq:eq2}
\end{equation}

\noindent
where $A_i$ and $\sigma_i$ are constants. The most difficult issue was the derivation of
these constants. Since AMICA has an approximate 10$^3$ dynamic range, it
is impossible to derive the faint widespread PSFs using stellar
images due to the faintness of total flux of stars. Instead, we used
images of Itokawa and the Moon to determine the broad PSFs.
The total intensity of these disks was bright enough to derive the
broad PSFs. First, we fixed $N$ and $\sigma_i$ for our convenience. We
assumed that $N$=6 and $\sigma_i$=$2^{i+2}$ when $N\leq$4 and that $\sigma_5$=110
and $\sigma_6$=710. There was no physical implication to choosing
these values. We chose these $\sigma_i$ values to fit
the wide spatial range 
of the attenuation function from the slightly large scale of a focused component
to the entire FOV. Next, we made a convolution image, $f_\mathrm{ufoc}(x,y)$
$\ast$ $I(x,y)$, where $r=\sqrt{x^2+y^2}$. $I(x,y)$ is the observed
intensity of Itokawa and the 
Moon in CCD coordinates assuming $A_i$ values arbitrarily. 
The image,
$f_\mathrm{ufoc}(x,y)$ $*$ $I(x,y)$, provides the evaluation of scattered
light for the entire image. The original pixel values, $I(x,y)$, were
subtracted using $f_\mathrm{ufoc}(x,y)$ $*$ $I(x,y)$, and we obtained the
image after the scattered light correction. 
We searched for the set of the $A_i$ values changing them
simultaneously. We set the criterion for fitting to suppress the
scattered light intensity at ~ 1\% level of Itokawa or lunar disk
brightness. We noticed
that small objects, i.e., Itokawa at the GP phase or the Moon, in the AMICA FOV were
useful in deriving $A_i$ for smaller $i$, whereas the large Itokawa images at the
HP phase was useful to derive $A_i$ for larger $i$. To derive the coefficient
$A_i$, we used images taken from May 16, 2004 to September 29, 2005. The
names of the images used are listed in Table~2. This includes the lunar and
Itokawa images with different apparent sizes on the detector.
We summarized the best fit of $A_i$ in Table~\ref{tab:Ai}.
In addition, we show the PSFs for all AMICA channels in Figure
~\ref{fig:f4}. As we suggested, PSF at longest channels (i.e. zs-band)
shows a wide profiles while PSFs  at b, v, and w-bands have narrow
profiles. Figure~\ref{fig:f5} shows the images after the scattered light
correction. Since we set the criterion for the fitting at  1\% level of
the object intensity, there still remains extended source in the sky region.
We found that the extended light is an artifact of AMICA (what is
called, a ghost), which is caused by a reflection from a surface near
the detector. Because we have no way to subtract the ghosts, we tolerate
the residuals below 1\% level of the object signal.

\begin{table}
 \footnotesize
 \caption{The coefficient $A_i$ ($10^{-4}$) and $\sigma_i$ (pixel) determined using the summation of Gaussian functions.
}
 \label{tab:Ai}
 \begin{center}
  \begin{tabular}{lccccccc}
   \hline
      & $A_1$ & $A_2$ & $A_3$ & $A_4$ & $A_5$ & $A_6$ \\
   \hline
   ul-band & 12.0 & 8.0 & 1.2 & 1.0 & 0.8 & 0.7 \\
   b-band  & 10.0 & 1.5 & 0.3 & 0.4 & 0.4 & 0.5 \\
   v-band  & 10.0 & 1.5 & 0.3 & 0.4 & 0.4 & 0.5 \\
   w-band  & 10.0 & 1.5 & 0.6 & 0.8 & 0.7 & 0.6 \\
   x-band  &  9.0 & 3.5 & 2.0 & 2.7 & 2.2 & 0.5 \\
   p-band  & 10.0 & 5.0 & 8.3 & 4.0 & 6.4 & 1.8 \\
   zs-band & 50.0 & 16.0& 6.0 & 9.0 & 9.5 & 4.5 \\
   \hline
   \hline
   &$\sigma_1$ &$\sigma_2$ &$\sigma_3$ &$\sigma_4$ &$\sigma_5$ &$\sigma_6$ \\
   \hline
   all-bands& 8 & 16 & 32 & 64 & 110 & 710 \\
   \hline
  \end{tabular}
 \end{center}
\end{table}

\renewcommand{\arraystretch}{0.8}
\begin{table}
 \footnotesize
  \caption{Data used for the determination of PSFs.}
\label{tab:first}
  \begin{center}
    \begin{tabular}{lllll}
      \hline
     Date & Object & File name & Filter & Distance (km) \\
      \hline

     2004-05-16 & Moon & ST\_1032725271 & ul & N/A\\
     2004-05-16 & Moon & ST\_1032730140 & b  & N/A\\
     2004-05-16 & Moon & ST\_1032735009 & v  & N/A\\
     2004-05-16 & Moon & ST\_1032739861 & w  & N/A\\
     2004-05-16 & Moon & ST\_1032744730 & x  & N/A\\
     2004-05-16 & Moon & ST\_1032749599 & zs & N/A\\
     2004-05-16 & Moon & ST\_1032754451 & p  & N/A\\

     \\

     2004-05-17 & Moon & ST\_1035434526 & ul & N/A\\
     2004-05-17 & Moon & ST\_1035439395 & b  & N/A\\
     2004-05-17 & Moon & ST\_1035444247 & v  & N/A\\
     2004-05-17 & Moon & ST\_1035449116 & w  & N/A\\
     2004-05-17 & Moon & ST\_1035453985 & x  & N/A\\
     2004-05-17 & Moon & ST\_1035458837 & zs & N/A\\
     2004-05-17 & Moon & ST\_1035463706 & p  & N/A\\

     \\

     2005-09-17 & Itokawa & ST\_2385540425 & ul & 16.3\\
     2005-09-17 & Itokawa & ST\_2385559680 & b  & 16.3\\
     2005-09-17 & Itokawa & ST\_2385578902 & v  & 16.3\\
     2005-09-17 & Itokawa & ST\_2385598109 & w  & 16.3\\
     2005-09-17 & Itokawa & ST\_2385617364 & x  & 16.3\\
     2005-09-17 & Itokawa & ST\_2385636586 & zs & 16.3\\
     2005-09-17 & Itokawa & ST\_2385655809 & p  & 16.3\\

     \\

     2005-09-25 & Itokawa & ST\_2406708286 & ul & 13.9\\
     2005-09-25 & Itokawa & ST\_2406717897 & b  & 13.9\\
     2005-09-25 & Itokawa & ST\_2406727508 & v  & 13.9\\
     2005-09-25 & Itokawa & ST\_2406737168 & w  & 13.9\\
     2005-09-25 & Itokawa & ST\_2406746796 & x  & 13.9\\
     2005-09-25 & Itokawa & ST\_2406756407 & zs & 13.9\\
     2005-09-25 & Itokawa & ST\_2406766067 & p  & 13.9\\

     \\

     2005-09-25 & Itokawa & ST\_2407399514 & ul & 13.2\\
     2005-09-25 & Itokawa & ST\_2407409109 & b  & 13.2\\
     2005-09-25 & Itokawa & ST\_2407418769 & v  & 13.2\\
     2005-09-25 & Itokawa & ST\_2407428397 & w  & 13.2\\
     2005-09-25 & Itokawa & ST\_2407438008 & x  & 13.2\\
     2005-09-25 & Itokawa & ST\_2407447668 & zs & 13.2\\
     2005-09-25 & Itokawa & ST\_2407457296 & p  & 13.2\\

     \\

     2005-09-29 & Itokawa & ST\_2418659460 & b  & 7.5\\
     2005-09-29 & Itokawa & ST\_2418699769 & v  & 7.5\\
     2005-09-29 & Itokawa & ST\_2418768895 & w  & 7.5\\
     2005-09-29 & Itokawa & ST\_2418807291 & p  & 7.5\\

     \\

     2005-10-17 & Itokawa & ST\_2468169379 & ul & 7.4\\
     2005-10-17 & Itokawa & ST\_2468172304 & b  & 7.4\\
     2005-10-17 & Itokawa & ST\_2468175197 & v  & 7.6\\
     2005-10-17 & Itokawa & ST\_2468178122 & w  & 7.6\\
     2005-10-17 & Itokawa & ST\_2468181047 & x  & 7.6\\
     2005-10-17 & Itokawa & ST\_2468183940 & zs & 7.4\\
     2005-10-17 & Itokawa & ST\_2468186849 & p  & 7.4\\


     \hline
    \end{tabular}
  \end{center}
\end{table}
\renewcommand{\arraystretch}{1}

\begin{figure}
  \begin{center}
  \includegraphics[width=80mm]{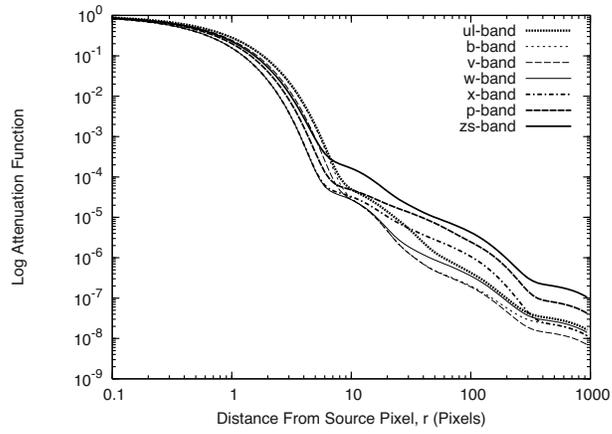}
  \caption{PSFs for all AMICA channels.}
\label{fig:f4}
  \end{center}
\end{figure}

\begin{figure}
 \begin{center}
  \includegraphics[width=80mm]{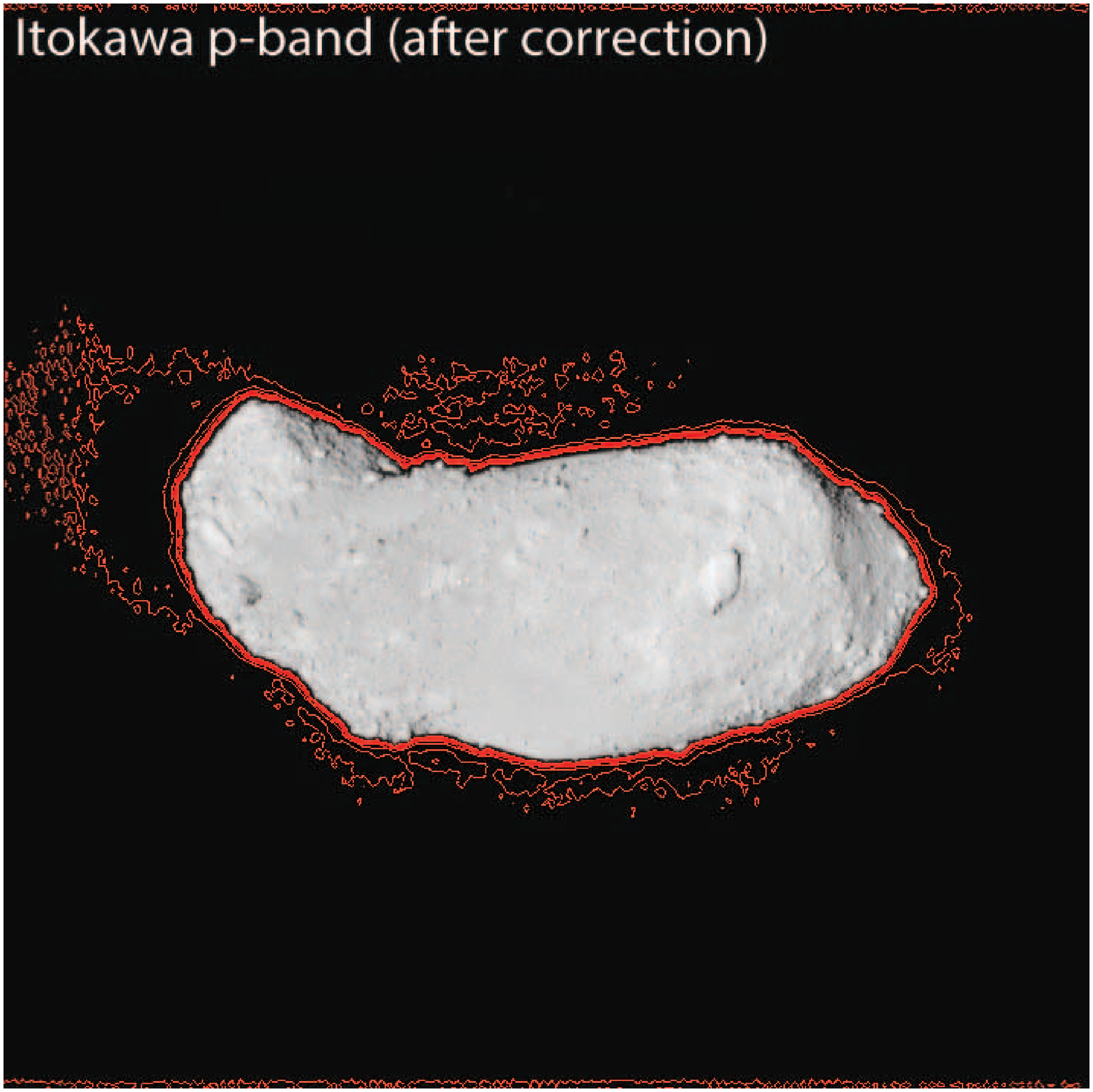}
  \includegraphics[width=80mm]{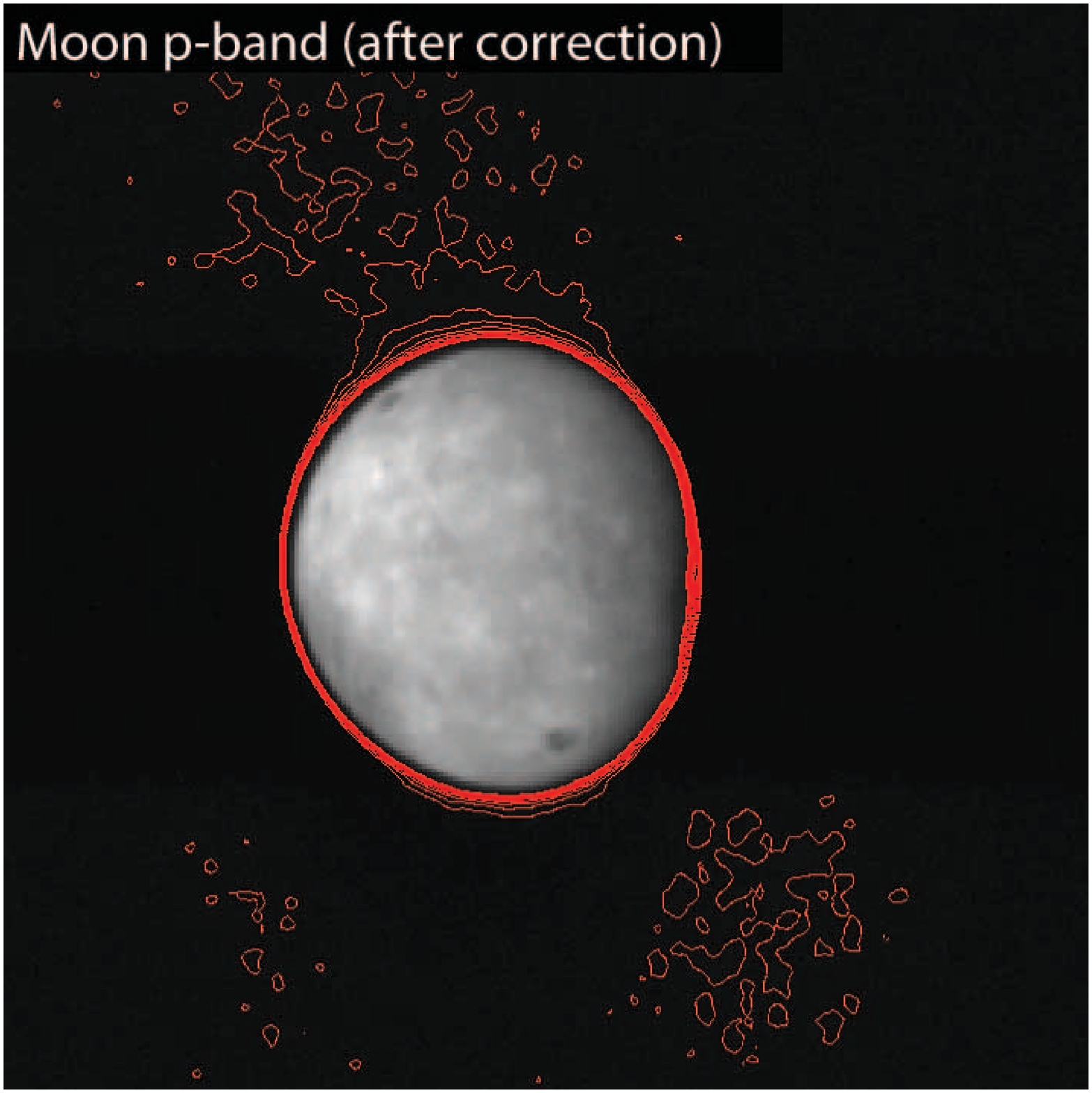}
  \caption{Images after the scattered light
  correction. The original data are the same as figure \ref{fig:f1}. The
  intensity of the residual is at the 1\% level of Itokawa or moon disk.
}\label{fig:f5} 
 \end{center}
\end{figure}

\subsection{Evaluation of the Correction Effect}

To evaluate how well the subtraction technique with the broad PSFs,
$f_\mathrm{ufoc}(r)$, suppresses the dim halo brightness, we used
another set of multiband images taken on October 17, 2005. Figures~\ref{fig:f6} and \ref{fig:f7}
show an example result
taken on October 17, 2005. Figure~\ref{fig:f6} is an image of Itokawa's
polar region where the shadows from the other parts of Itokawa were cast.
In the image, there was a large-scale dim halo surrounding the entire
disk of Itokawa before the scattered light correction in near-infrared
channels. Since lights from these shadow areas are believed to be zero,
we selected this set of images as our benchmark for our
correction method. The solid lines in Figure~\ref{fig:f7} are the
p-band intensity profiles along two lines from ``A'' to ``B'' and from
``C'' to ``D'' in Figure~\ref{fig:f6}. In the
shadow areas marked S1, S2 and S3, the intensity in the
uncorrected data are significantly higher, i.e., $>$15\% of Itokawa's disk
intensity, than the zero intensity level because the light is smeared out from
the nearby pixels due to the scattered light inside the AMICA optics. We
subtracted the scattered light component using the above technique, and
found that the intensity of small shadows, marked ``S1--S3'' in Figure~\ref{fig:f6},
and limbs are 0$\pm$1\% of Itokawa disk intensity.  It
may be reasonable to think that the shadow areas were irradiated by
another sunlit Itokawa surface. However, we believe the effect of this was
too small to be detected. If the effect were noticeable, then there should be a
discontinuous boundary at the LIMB position in Figure~\ref{fig:f7},
which corresponds to the limb of the Itokawa's disk in the shadowed area.
Thanks to our technique, the scattered light in small scale shadowed
areas (S1--3 in Figure~\ref{fig:f7} left), large scale shadowed area (at
the distance of 300--450 pixel in Figure~\ref{fig:f7}), and sky region
are corrected to an accuracy of $\sim$1\% of the Itokawa
disk. Therefore, our subtraction method works for the wide spatial range
of scattered light. 


\begin{figure}
  \begin{center}
  \includegraphics[width=80mm]{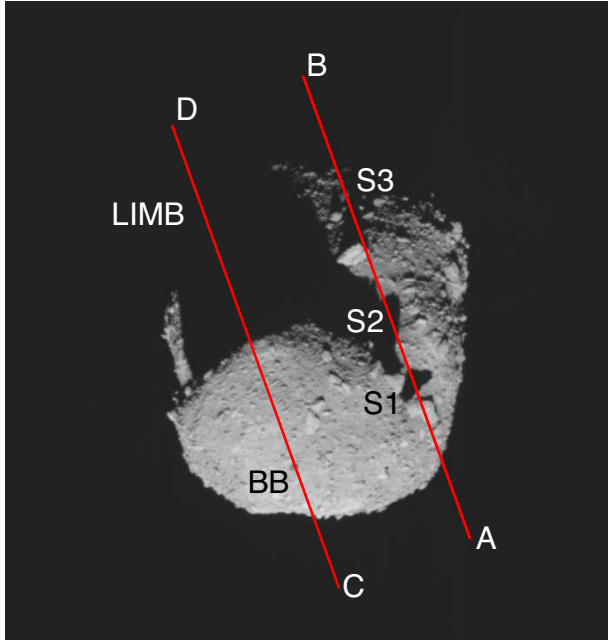}
  \caption{v-band image of Itokawa taken on October 17, 2005. This image
   was taken without binning in lossy compression mode. The cut profiles
   of the p-band lossless data were shown in Figure~\ref{fig:f7} along the
   line between ``A''--``B'' and ``C''--''D''. We show three prominent
   shadows of boulders projected on the Itokawa surface by ``S1'', ``S2'', and
   ``S3''. ``BB'' indicates the position of a dark boulder called Black
   Boulder. ``LIMB'' shows a position of Itokawa's limb in the
   shadow.}\label{fig:f6}
  \end{center}
\end{figure}

\begin{figure}
  \begin{center}
   \includegraphics[width=80mm]{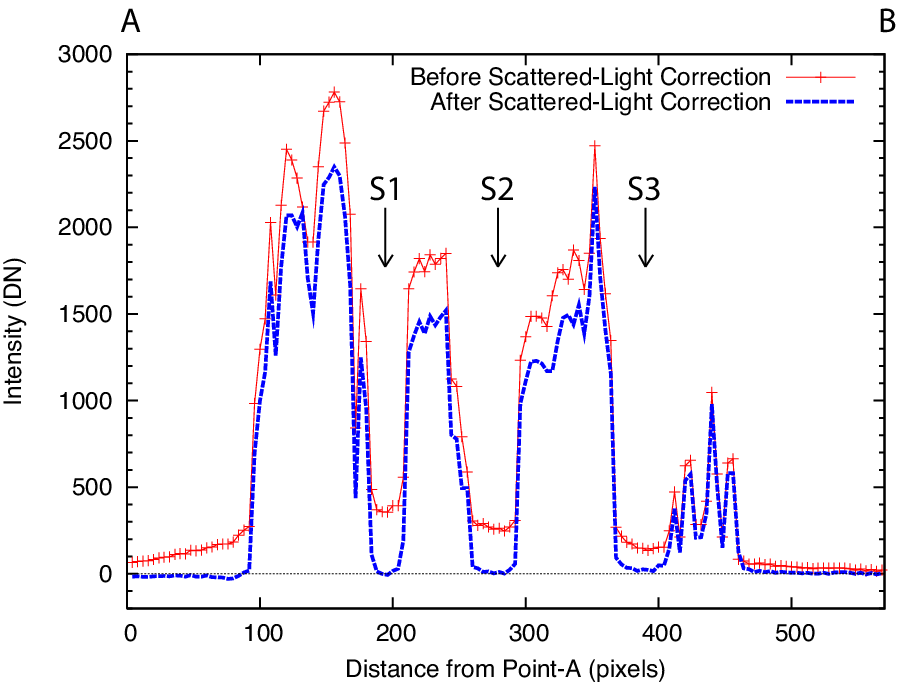}
   \includegraphics[width=80mm]{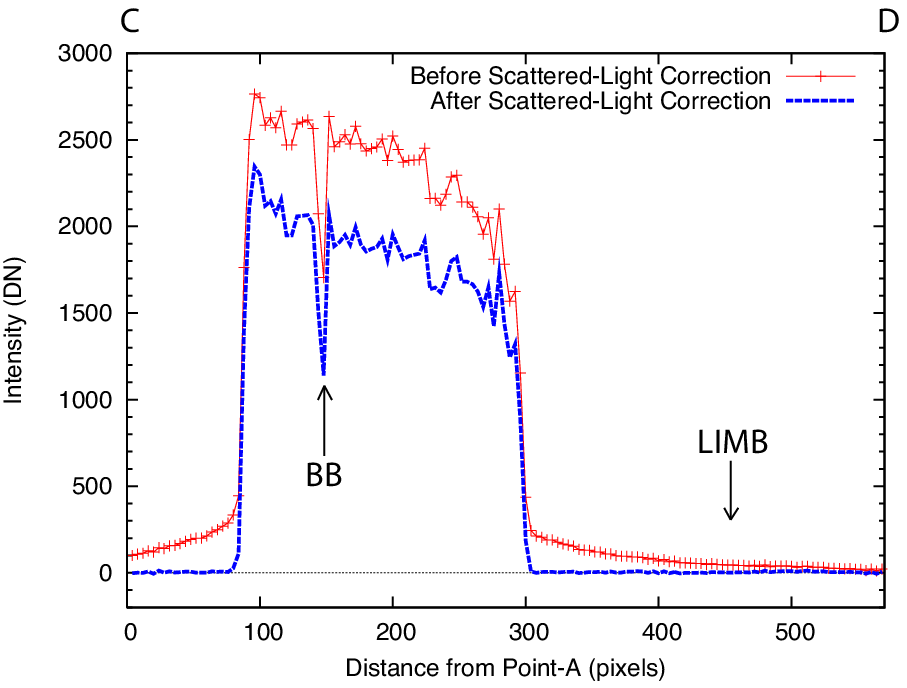}
  \caption{Cut profiles along the line in Figure~\ref{fig:f6} using
   scattered-light corrected and uncorrected data. Three labels,
   S1--S3, correspond to shadow regions on Itokawa, whose data counts
   should be zero without scattered light contamination.}\label{fig:f7}
  \end{center}
\end{figure}

We compared the AMICA PSFs with those of Galileo SSI. We have no
experience in deriving the wide range of PSFs for the spacecraft's camera, so
we took a heuristic approach. It is important to compare the AMICA
PSFs with those from the previous mission's onboard camera. \citet{Gaddis1995}
provided the PSFs of the Galileo onboard camera, SSI, for two
different filters. In the paper, they reported that the largest
amount of attenuation due to the scattered light was observed in the 1MC
(990 nm) filter data and the least amount was observed in the GRN (560~nm).
We selected the PSFs of AMICA's two filters whose central wavelengths
are closest to those of these SSI filters.
Figure~\ref{fig:f8} shows the
comparison of the PSFs between the Hayabusa/AMICA and Galileo SSI. In these
plots, we found that the general trends of the AMICA PSFs were similar to
those of the SSI. They exhibited central focused components that extended up to 2--10
pixels. The full width at half maximum (FWHM) of the SSI GRN is better than that of the AMICA v-band, thus
implying that the SSI GRN obtained sharper images than AMICA. The other
parts, especially the broad PSFs, which are of great concern in this
research, were quite similar to one another. Beyond $\sim$10 pixels, the
PSFs attenuate 10$^{-7}$ at $r$=100
pixels with respect to flux near the center (at $r$=0.1 pixel) around
553--560 nm, while 10$^{-6}$ level at $r$=100 pixels around 960--990~nm.
Thus, there appears only a one order of magnitude difference in PSFs between the
two wavelength results, which causes big differences in the scattered light.

\begin{figure}
  \begin{center}
   \includegraphics[width=80mm]{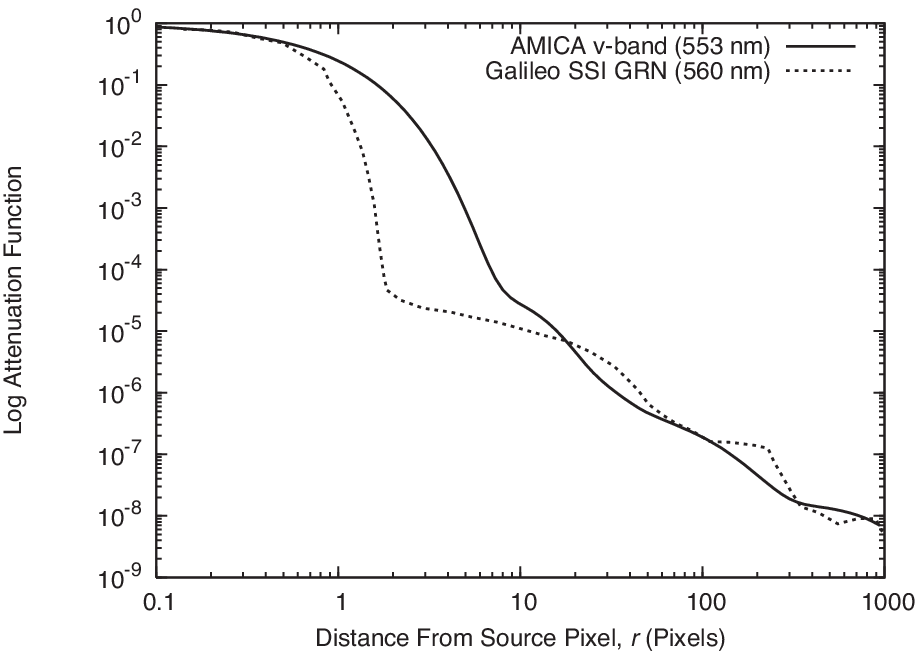}
   \includegraphics[width=80mm]{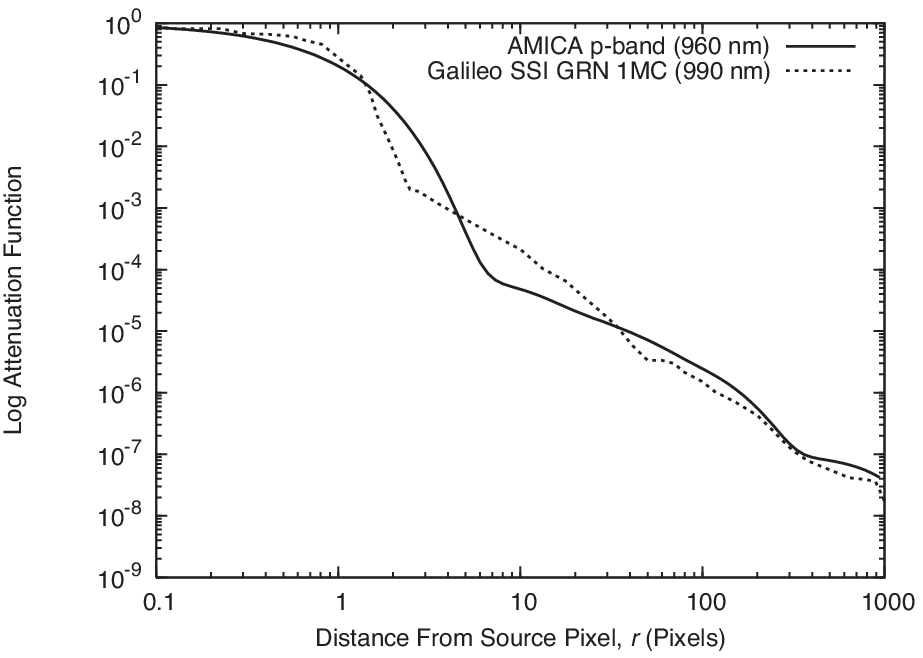}
  \caption{Plots of the PSFs for two AMICA filters, i.e., v-
   and p-bands. For comparison, we show the PSFs of the Galileo spacecraft
   onboard camera, SSI, which contains two filters.}\label{fig:f8}
  \end{center}
\end{figure}

Finally, we estimated the remaining uncertainty in Itokawa relative
reflectance. The data was acquired with a substantial signal-to-noise
ratio. Considering the readout noise of 60 e$^-$, dark accumulation rate
of 0.14 DN s$^{-1}$ pixel$^{-1}$, gain factor of 17e$^-$ DN$^{-1}$ and
typical Itokawa intensity of 3000 DN \citep{Ishiguro2010}, we estimated
the S/N$>$200. On the other hand, 
\citet{Ishiguro2010} verified that the accuracy of the disk-integrated
relative reflectance was about 1\%, the flat-field uncertainty was
$<$3\%, although the scattered light correction remained unsettled. Our
new data reduction technique could suppress the scattered light
intensity at 1\% level of Itokawa intensity, which is negligible
compared with the flat-field uncertainty. Consequently, the overall data
reduction scheme including the scattered light correction enables to
recognize ~3\% regional differences in the relative reflectance spectra
of Itokawa. 

\section{Results}

After subtracting the scattered light component, we verified the
produced maps of $R_w/R_b$ and $R_p/R_w$ using a set of data at the GP phase,
where $R_b$, $R_w$, and $R_p$ represent the reflectance in the b-, w- and p-band,
respectively, as well as the after bias, readout smear, flat-fielding and scattered
light correction, and flux calibration (Figure~\ref{fig:color_map}).
We used the following images for the analysis: ST\_2385559680 (b-band), ST\_2385598109 (w-band) and
ST\_2385655809 (p-band). In these images, the sampling
site of the Hayabusa spacecraft is visible near the center. In the map of
$R_w$/$R_b$  (Figure~\ref{fig:color_map} left), the brighter parts correspond
to regions of redder spectra, whereas the darker parts correspond to regions
of bluer spectra in the wavelength regions between 429 nm (b-band) and
700 nm (w-band). Conversely, in the map of $R_p/R_w$ (Figure~\ref{fig:color_map} right),
the darker parts correspond to the regions of
deeper absorption, while the redder parts correspond to regions of shallower
absorption around 960 nm. It is clear that the redder parts exhibit a shallower
absorption around 960 nm, while the bluer parts exhibit deeper absorption. This
trend can be explained by space weathering effects \citep{Chapman2004}.

\begin{figure}
  \begin{center}
   \includegraphics[width=80mm]{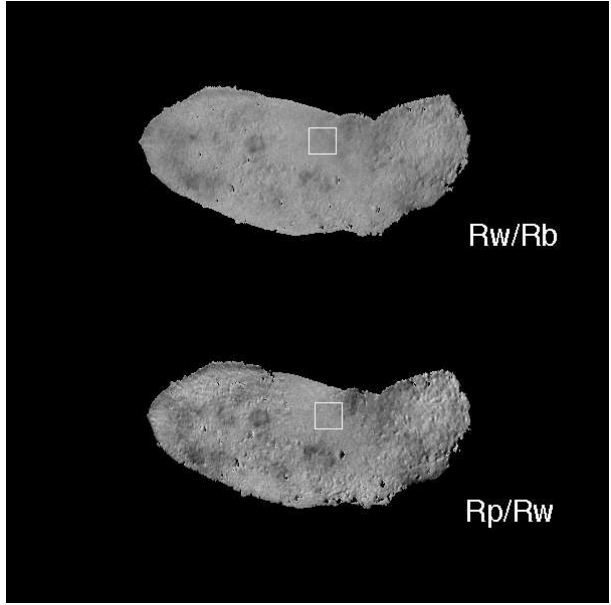}
  \caption{Maps of the $R_w/R_b$ (top) and $R_p/R_w$ (bottom), where $R_b$,
   $R_w$, and $R_p$ represent the reflectance in the b-, w-, and p-band,
   respectively. The position of the Hayabusa spacecraft
   landing site is indicated by a square based on the description in
   \citep{Yano2006}.}\label{fig:color_map}
  \end{center}
\end{figure}

Figure~\ref{fig:correlation} shows the correlation between $R_w$/$R_b$
(horizontal axes) and $R_p/R_w$ (vertical axes). For comparison, we show
the correlation using images before the scattered light correction
(Figure~\ref{fig:correlation} left). The trend of space weathering becomes
more pronounced after the scattered light correction. We derived
a correlation coefficient of 0.71 and 0.83 using the images generated before and after the scattered light
correction, respectively.
In Figure~\ref{fig:correlation} right, some terrains exhibit data points
at the lower left, thereby suggesting that fresh materials were excavated. The
square was obtained at a rim of the Komaba crater
\citep{Demura2006,Hirata2009}. The sampling site in the MUSES-SEA terrain
shows that $R_w/R_b$=1.38$\pm$0.01 and $R_p/R_w$=0.89$\pm$0.01, which is
nearly the average value of the entire Itokawa disk, i.e., $R_w/R_b$=1.383 and
$R_p/R_w$=0.880.

\begin{figure}
  \begin{center}
   \includegraphics[width=80mm]{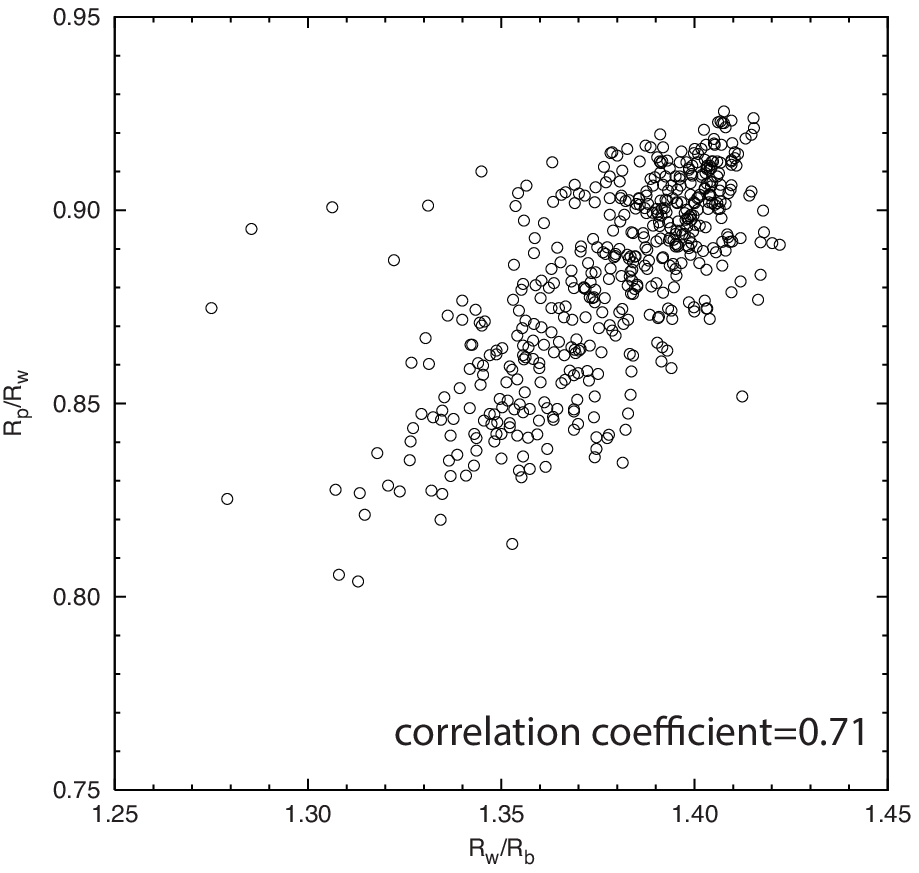}
   \includegraphics[width=80mm]{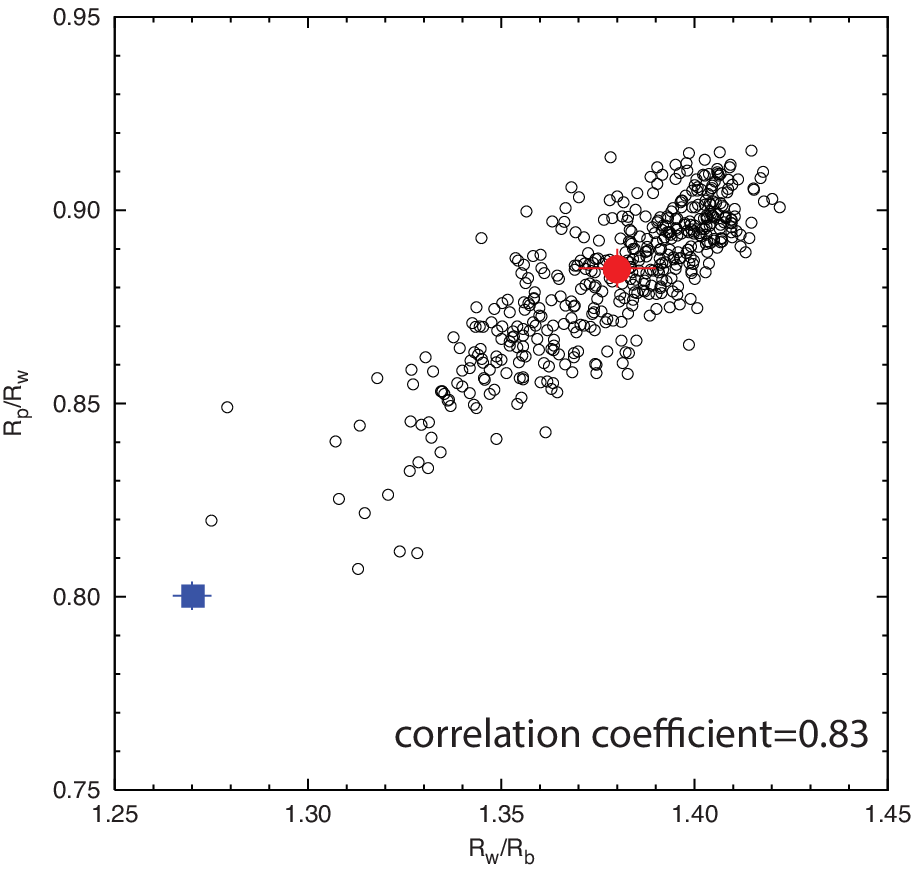}
  \caption{Correlation between $R_w$/$R_b$ and $R_p/R_w$ using data
   before (left) and after (right) the scattered light correction, respectively.
   The diagonal trend from the lower left to upper
   right can be explained by space weathering \citep{Chapman2004}. We
   derived a correlation coefficient of 0.71 and 0.83 using the scattered light
   uncorrected and corrected data, respectively. The
   square and filled circles denote the values at the Komaba crater and
   Hayabusa sampling site, respectively.}\label{fig:correlation}
  \end{center}
\end{figure}


In 2010, the Hayabusa mission was brought to a successful conclusion.
The remote-sensing
observations revealed that the small asteroid had a rubble pile structure
\citep{Fujiwara2006} and the surface materials experienced space weathering
\citep{Hiroi2006}. Together with a sample analysis, the Hayabusa mission
achieved the first connection
between ordinary chondrite
meteorites and S-type asteroids \citep{Noguchi2011}. We expect our new
method will provide further knowledge on the global and regional diversity
of the space weathering effect and promote a better understanding of the
evolution process of Itokawa on both the microscopic and macroscopic scales.


\bigskip

The detailed inflight testing of the AMICA was made possible by the
dedication and hard work of many members of the Hayabusa Mission
teams. The author thank AMICA PI, Dr. J. Saito, and Profs. T. Hashimoto,
and T. Kubota for their effort to the development and the operation of
ONC/AMICA. The author has been encouraged by Hayabusa 2 in preparation
for the upcoming space mission. This paper was written
at UCLA with a support of Prof. David Jewitt. The image processing
method was developed at Seoul National University by the National
Research Foundation of Korea (NRF) grant funded by the Korean government
(MEST) (No. 2012R1A4A1028713).



\end{document}